\documentstyle{mn}

\newif\ifAMStwofonts

\ifoldfss
  \ifCUPmtlplainloaded \else
    \NewTextAlphabet{textbfit} {cmbxti10} {}
    \NewTextAlphabet{textbfss} {cmssbx10} {}
    \NewMathAlphabet{mathbfit} {cmbxti10} {} 
    \NewMathAlphabet{mathbfss} {cmssbx10} {} 
  \fi
  \ifAMStwofonts
    \ifCUPmtlplainloaded \else
      \NewSymbolFont{upmath} {eurm10}
      \NewSymbolFont{AMSa} {msam10}
      \NewMathSymbol{\upi}     {0}{upmath}{19}
      \NewMathSymbol{\umu}     {0}{upmath}{16}
      \NewMathSymbol{\upartial}{0}{upmath}{40}
      \NewMathSymbol{\leqslant}{3}{AMSa}{36}
      \NewMathSymbol{\geqslant}{3}{AMSa}{3E}

       \let\le=\leqslant
       \let\ge=\geqslant
    \fi
  \fi
\fi 

\ifnfssone
  \newmathalphabet{\mathit}
  \addtoversion{normal}{\mathit}{cmr}{m}{it}
  \addtoversion{bold}{\mathit}{cmr}{bx}{it}
  \newmathalphabet{\mathbfit} 
  \addtoversion{normal}{\mathbfit}{cmr}{bx}{it}
  \addtoversion{bold}{\mathbfit}{cmr}{bx}{it}
  \newmathalphabet{\mathbfss} 
  \addtoversion{normal}{\mathbfss}{cmss}{bx}{n}
  \addtoversion{bold}{\mathbfss}{cmss}{bx}{n}
  \ifAMStwofonts
    \ifCUPmtlplainloaded \else
      \UseAMStwoboldmath
      \makeatletter
      \new@mathgroup\upmath@group
      \define@mathgroup\mv@normal\upmath@group{eur}{m}{n}
      \define@mathgroup\mv@bold\upmath@group{eur}{b}{n}
      \edef\UPM{\hexnumber\upmath@group}
      \new@mathgroup\amsa@group
      \define@mathgroup\mv@normal\amsa@group{msa}{m}{n}
      \define@mathgroup\mv@bold\amsa@group{msa}{m}{n}
      \edef\AMSa{\hexnumber\amsa@group}
      \makeatother
      \mathchardef\upi="0\UPM19
      \mathchardef\umu="0\UPM16
      \mathchardef\upartial="0\UPM40
      \mathchardef\leqslant="3\AMSa36
      \mathchardef\geqslant="3\AMSa3E

       \let\le=\leqslant
       \let\ge=\geqslant
    \fi
  \fi
\fi 

\ifnfsstwo
  \DeclareMathAlphabet{\mathbfit}{OT1}{cmr}{bx}{it}
  \SetMathAlphabet\mathbfit{bold}{OT1}{cmr}{bx}{it}
  \DeclareMathAlphabet{\mathbfss}{OT1}{cmss}{bx}{n}
  \SetMathAlphabet\mathbfss{bold}{OT1}{cmss}{bx}{n}
  \ifAMStwofonts
    \ifCUPmtlplainloaded \else
      \DeclareSymbolFont{UPM}{U}{eur}{m}{n}
      \SetSymbolFont{UPM}{bold}{U}{eur}{b}{n}
      \DeclareSymbolFont{AMSa}{U}{msa}{m}{n}
      \DeclareMathSymbol{\upi}{0}{UPM}{"19}
      \DeclareMathSymbol{\umu}{0}{UPM}{"16}
      \DeclareMathSymbol{\upartial}{0}{UPM}{"40}
      \DeclareMathSymbol{\leqslant}{3}{AMSa}{"36}
      \DeclareMathSymbol{\geqslant}{3}{AMSa}{"3E}

       \let\le=\leqslant
       \let\ge=\geqslant
    \fi
  \fi
\fi 

\ifCUPmtlplainloaded \else
  \ifAMStwofonts \else 
    \def\upi{\pi}
    \def\umu{\mu}
    \def\upartial{\partial}
  \fi
\fi

\title[Long-Period Orbit of V630 Cas]{The Long-Period 
Orbit of the Dwarf Nova V630 
Cassiopaeia\protect\thanks{Based on observations collected at the 
William Herschel Telescope 
operated on the island of La Palma by the Isaac Newton 
Group in the Spanish Observatorio del Roque de los
Muchachos of the Instituto de Astrofisica de Canarias}}
\author[Orosz et al.]
       {Jerome A. Orosz$^1$\thanks{Visiting Astronomer at Kitt Peak National 
Observatory (KPNO), which is operated by AURA, Inc., under a cooperative 
agreement 
with the National Science Foundation.}, 
John R. Thorstensen$^2$, R. Kent Honeycutt$^3$\\
	$^1$Astronomical Institute, Utrecht University, 3508 TA Utrecht,
        The Netherlands \\
	$^2$Department of Physics and Astronomy,
	6127 Wilder Laboratory, Dartmouth College,
	Hanover, NH 03755, USA \\
        $^3$Astronomy Department, Indiana University, Bloomington, IN
        47405, USA}
\date{Received:}

\pagerange{\pageref{firstpage}--\pageref{lastpage}}

\begin{document}

\maketitle

\label{firstpage}

\begin{abstract}
We present extensive spectroscopy and photometry
of the dwarf nova V630 Cassiopeiae.  A late-type (K4-5)
absorption spectrum is easily detectable, from which we derive
the orbital parameters.  We find a spectroscopic period of
$P=2.56387\pm 4\times 10^{-5}$ days and a semiamplitude of
$K_2=132.9\pm 4.0$ km s$^{-1}$.  The resulting mass function, which 
is a firm lower limit on the mass of the white dwarf, is then
$f(M)=0.624\pm 0.056\,M_{\odot}$.  The secondary star is a ``stripped
giant'', and using relations between the core mass and the luminosity
and the core mass and the radius we derive a lower limit of 
$M_2\ge 0.165\,M_{\odot}$ for the secondary star.  The rotational
velocity of the secondary star is not resolved in our spectra and
we place a limit of $V_{\rm rot}\sin i< 40$ km s$^{-1}$.
The long-term light curve shows variations
of up to 0.4 mag on short (1-5 days) time scales, and variations of
0.2-0.4 mag on longer (3-9 months) time scales.  In spite of these variations,
the ellipsoidal light curve of the secondary star is easily seen when
the data are folded on the spectroscopic ephemeris.  Ellipsoidal models
fit to the mean light curve give an inclination in the range
$66.96 \le i \le 78.08$ degrees (90 per cent confidence).
This inclination range, and the requirement that $M_2\ge 0.165\,M_{\odot}$
and $V_{\rm rot}\sin i< 40$ km s$^{-1}$ yields
a white dwarf mass of $M_1=0.977^{+0.168}_{-0.098}\,M_{\odot}$
and a secondary star mass of $M_2=0.172^{+0.029}_{-0.012}\,M_{\odot}$
(90 per cent confidence limits).
Our findings confirm the suggestion of Warner (1994), namely that
V630 Cas is rare example of a dwarf nova with a long orbital period.
\end{abstract}


\section{Introduction}

V630 Cassiopeiae is a poorly studied cataclysmic variable star.
This star had a 4.8 mag in outburst in 1950 \cite{whi73}, 
and a 2 mag outburst
observed in late 1992 \cite{hon93}.  The optical spectrum in
quiescence shows absorption lines indicative of a late K-star
and low-level
excitation emission lines \cite{szk92}.
V630 Cas is classified in the Duerbeck (1987) catalog as a likely 
WZ Sagittae-type dwarf nova, a particular subtype of
outbursting cataclysmic variable stars which have relatively large
amplitude outbursts and relatively long intervals between outbursts.
However, based on its similarity to BV Cen, GK Per, and V1017 Sgr,
Warner (1994) suggested that V630 Cas is a rare example of a dwarf
nova with a long orbital period.  Warner (1994) predicted a period
of $\approx 6$ days, based on a correlation between the orbital period and
the recurrence time between outbursts, and on
a correlation between
the orbital period and the rate of rise to outburst.  We have collected
spectroscopic observations of the source over a number of years, and from
these we show that the orbital period is indeed relatively long with
$P=2.56387\pm 0.00004$ days. We also present a long-term $V$-band light
curve of the source.  Our observations allow us to derive reasonably precise
component masses and other interesting binary parameters.  We discuss
our observations and analysis, and the implications of these below.

\section{Observations}

\subsection{Photometry}

\begin{figure}
\vspace{7.2cm}
\includegraphics{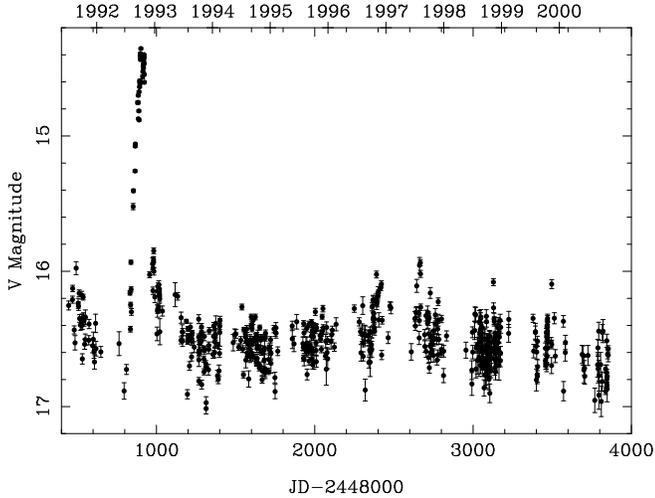} 
\caption{The Roboscope $V$-band light curve of V630 Cas, complete until
November, 2000.
}
\label{robo}
\end{figure}

This V630 Cas photometry was obtained by RoboScope \cite{hon92a}, 
an unattended 0.41-m telescope in central Indiana that is devoted to
long-term photometric monitoring of accretion systems.  The data were
reduced using the method of incomplete ensemble photometry \cite{hon92b}.
The ensemble solution used 76 comparison stars and 579 CCD exposures 
obtained over the interval 1991
July to 2000 November; the zero-point was
established
using 12 secondary standards from Henden \& Honeycutt (1995).  
Fig.\ \ref{robo}
shows the complete light curve, including an extended 
3-month long outburst
in 1992.  This 2 mag outburst was described in Honeycutt et al.\ (1993),
and further discussed in Warner (1994).  We also see in Fig.\
\ref{robo} a rise
of $\sim$0.2 mag over the four years following the outburst.  This is
followed by a similar $\sim$0.2 mag decline during the succeeding four years.
Superimposed on this 1993-96 rise and 1997-2000 decline is an additional 
$\sim$0.2 mag of scatter that substantially exceeds the observational
error.  
There are changes of up to 0.4 mag
that are unresolved at our typical data spacing of 1 to 5 days,
superimposed on resolved 0.2-0.4 mag changes with time scales of 3 to 9 
months.

\begin{table*}
  \caption{Journal of Spectroscopy}
  \begin{tabular}{ccccc}
\hline
Date (UT) & $N$ & UT start & UT end & telescope \\
\hline
1993-12-08 & 2 &  6:49 & 7:03 & KPNO 4m \\
1993-12-10 & 18 & 1:21 & 7:17 & KPNO 4m \\
1994-11-05 & 6 & 2:53 & 3:50 & KPNO 4m \\
1994-11-06 & 11 & 1:47 & 9:24 & KPNO 4m \\
1994-11-07 & 11 & 1:44 & 3:50 & KPNO 4m \\
1995-10-01 & 3 & 2:51 & 4:35 & WIYN 3.5m \\
1995-10-02 & 2 & 4:58 & 5:22 & WIYN 3.5m \\
1995-10-03 & 2 & 2:56 & 3:33 & WIYN 3.5m \\
1997-09-19 & 2 & 5:41 & 10:46 & MDM 2.4m \\
1997-09-20 & 2 & 4:37 & 11:54 & MDM 2.4m \\
1997-09-21 & 1 & 9:19 & -- & MDM 2.4m \\
1997-09-22 & 1 & 8:55 & -- & MDM 2.4m \\
1997-09-23 & 2 & 2:37 & 9:20 & MDM 2.4m \\
1998-01-28 & 1 & 4:13 & -- & MDM 2.4m \\
1998-01-31 & 1 & 2:25 & -- & MDM 2.4m \\
1998-02-03 & 1 & 3:05 & -- & MDM 2.4m \\
1999-06-13 & 1 & 10:43 & -- & MDM 2.4m \\
1999-10-19 & 1 & 9:24 & -- & MDM 2.4m \\
1999-10-21 & 1 & 9:46 & -- & MDM 2.4m \\
1999-10-22 & 1 & 9:01 & -- & MDM 2.4m \\
1999-11-06 & 1 & 1:55 & -- & WHT 4.2m \\
1999-12-21 & 1 & 20:10 & -- & WHT 4.2m \\
2000-01-06 & 1 & 3:17 & -- & MDM 2.4m \\
2000-01-07 & 1 & 4:34 & -- & MDM 2.4m \\
2000-01-08 & 1 & 4:51 & -- & MDM 2.4m \\
\hline
\end{tabular}

\medskip

$N$ is the number of velocities included in the
analysis; for the MDM data, each velocity represents
the sum of several exposures, typically of 8 min. duration.
Times given are heliocentric times of mid-exposure of the
first and last velocity each night.
\label{tab1}
\end{table*}

\begin{figure}
\vspace{6.5cm}
\includegraphics{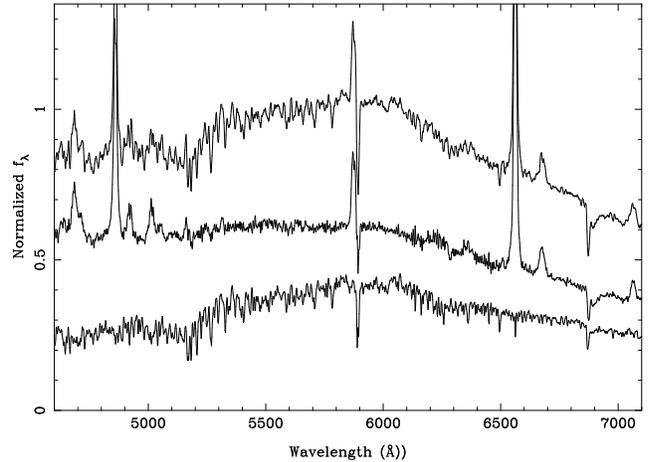} 
\caption{Top:  The averaged spectrum (in the rest frame of the secondary)
from the 1994 KPNO observations, normalised to unity at 5500~\AA. 
Bottom:  The spectrum of
HD 333388 (a K5 dwarf) observed with the same instrumentation.  The spectrum
was normalised to unity at 5500~\AA\ and scaled by 0.38.
Middle:  The difference spectrum, which is essentially the spectrum
of the accretion disc.
}
\label{spect}
\end{figure}

\begin{figure}
\vspace{8.5cm}
\includegraphics{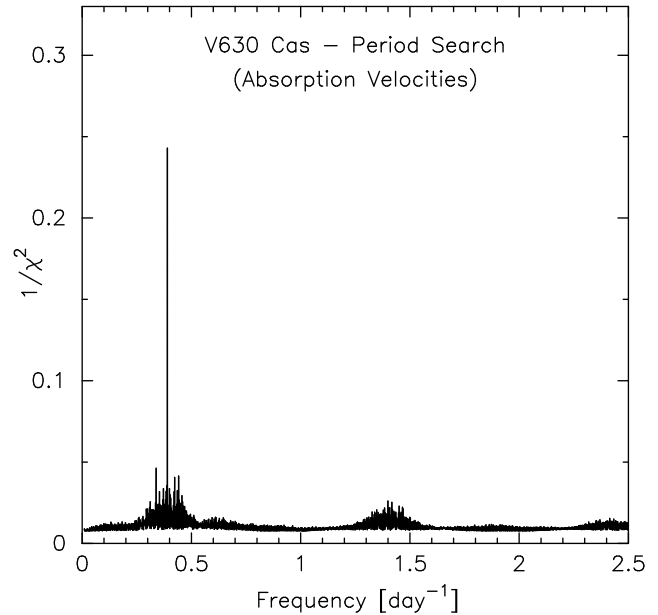} 
\caption{The results of the period search on the absorption line
velocities.  
}
\label{rgram}
\end{figure}

\begin{figure}
\vspace{6.5cm}
\includegraphics{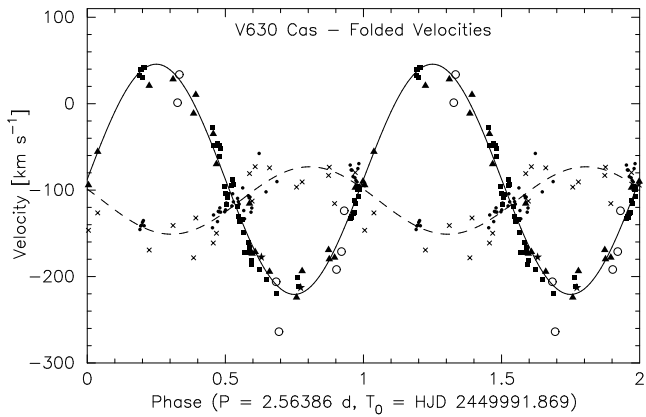} 
\caption{The phased absorption line and emission line velocities and
the corresponding best-fitting sinusoids.  The plot symbols are
filled squares:  KPNO 4-meter absorption line;
open circles:  WIYN Hydra  absorption line;
filled triangles: MDM 2.4m absorption line; 
5-pointed star:   WHT absorption line;
small round dots: KPNO 4-meter H-alpha emission velocities; and
small x's: MDM H-alpha emission velocities.
}
\label{foldRV}
\end{figure}

\subsection{Spectroscopy}

Table \ref{tab1} is a journal of the spectroscopic
observations.  The KPNO 4m spectra
were taken with the R/C spectrograph and a CCD detector, and
covered $\lambda\lambda$ 4450 -- 6930 with $\sim 3$ \AA\ resolution;
individual exposures were typically 600 s.  At the MDM Hiltner
2.4m telescope we used the modular spectrograph and a grating
covering 4300 -- 7400 \AA\ with 3 \AA\ resolution, but
with considerable vignetting toward the ends.  The WIYN
spectra were from the Hydra multi-fibre instrument; these
had inferior signal-to-noise and resolution, but proved useful
in confirming the ephemeris (below).  
The spectra from the 4.2m William Hershel Telescope (WHT) were
obtained through the service programme of the Isaac Newton Group of
telescopes on La Palma.  The instrumental configuration consisted of the
red arm of ISIS and the R1200R grating, yielding $\approx 1$~\AA\
resolution near H$\alpha$.  The two spectra each consist of the median of
three consecutive 20-minute exposures.
We were
careful to maintain an accurate wavelength calibration for
all the observations.  We reduced the raw CCD pictures to
one-dimensional, wavelength-binned, flux-calibrated spectra
using standard techniques implemented in 
IRAF\footnote{IRAF is distributed by the National Optical Astronomy
Observatories, which are operated by the Association of University for
Research in Astronomy, Inc., under cooperative agreement with the National
Science Foundation}.

\section{Analysis}

\subsection{Spectroscopic Period}

Our spectra of V630 Cas show Balmer emission lines, emission lines
of helium (most notably the lines at $\lambda 5876$ and
$\lambda 6678$), and relatively strong absorption lines from a $\approx$K
type star (see Fig.\ \ref{spect}).  
Similar features were also noted by Szkody \& Howell (1992). 
To measure the absorption spectrum radial velocities, we
obtained spectra of K-type stars using the same equipment, and
cross-correlated these with the V630 Cas spectra using the
IRAF tasks fxcor \cite{ton79} and xcsao \cite{kur98}.
The cross-correlations mostly yielded formal error estimates
$< 10$ km s$^{-1}$, but the mutual agreement of the standard
stars suggests that systematic effects limited the
uncertainty of a single velocity to $\sim 15$ km s$^{-1}$.

We searched for periods in the absorption velocities by
fitting them with sinusoids over a sufficiently dense grid
of test frequencies, from 0.01 to over 4 cycle d$^{-1}$.
Fig.\ \ref{rgram} shows the result, plotted as the inverse of the
mean square residual of the fit.   This period-finding method should be
especially suitable for the present data, in which the periodic signal
is expected to be accurately sinusoidal and the modulation
is expected to greatly exceed the noise.  In this circumstance
a very large spike is expected at the correct frequency, because
the small scatter at that frequency results in a much tighter
fit than at incorrect frequencies.  This expectation is
borne out here -- a frequency near 0.39 cycle d$^{-1}$
stands out prominently,
reflecting a uniquely good fit near 2.56 d.  The single
sharp spike shows that the cycle count
is uniquely determined over the 6-year span of the observations.
Table \ref{specparm}
gives parameters of the best sinusoidal fit to the
observations, and the folded radial velocity curve is shown in
Fig.\ \ref{foldRV}.

The emission lines were generally single-peaked, with the exception of
the spectra from 1994 November 7, which had double-peaked profiles with
an average peak-to-peak separation of $410\pm 35$ km s$^{-1}$.
We measured the radial velocities of the H$\alpha$ emission lines using the
double-Gaussian technique of Shafter (1983).  The Gaussians had a full width
at half maximum of 8~\AA\ and the separation between them was 24~\AA.
The emission line
radial velocities are shown folded on the orbital period in
Fig.\ \ref{foldRV}.  There is a great deal of scatter, but the emission
line radial velocities are generally out of
phase with respect to the absorption line radial velocities.
The formal semi-amplitude of the emission line radial velocity curve
is $K_1=39$ km s$^{-1}$.   However, we would not rely on the emission
line radial velocity curve to derive any dynamical information given
the large scatter seen in the velocities.

\begin{table}
  \caption{Spectroscopic elements}
  \begin{tabular}{lr}
\hline
parameter & value \\
\hline
Period, absorption lines (days)    &  $2.56387\pm 4\times 10^{-5}$ \\
$K_2$ velocity, absorption lines (km s$^{-1}$)  &  $132.9\pm 4.0$ \\
$\gamma_2$ velocity, absorption lines (km s$^{-1}$)  &  $-87.3\pm 2.3$ \\
$T_0$, absorption lines (HJD 2,400,000+) & $49994.4340\pm   0.0087$ \\
       &      \\
Period, emission lines (days)   &  $2.56375\pm 1.2\times 10^{-4}$ \\
$K_1$ velocity, emission lines (km s$^{-1}$)  &  $39.1\pm 4.9$ \\
$\gamma_1$ velocity, emission lines (km s$^{-1}$)  &  $-111.4\pm 2.7$ \\
$T_0$, emission lines (HJD 2,400,000+) & $49662.566\pm   0.034$ \\
\hline
  \end{tabular}
\label{specparm}
\end{table}

\subsection{Phase Averaged Light Curve}

\begin{figure}
\vspace{9.0cm}
\includegraphics{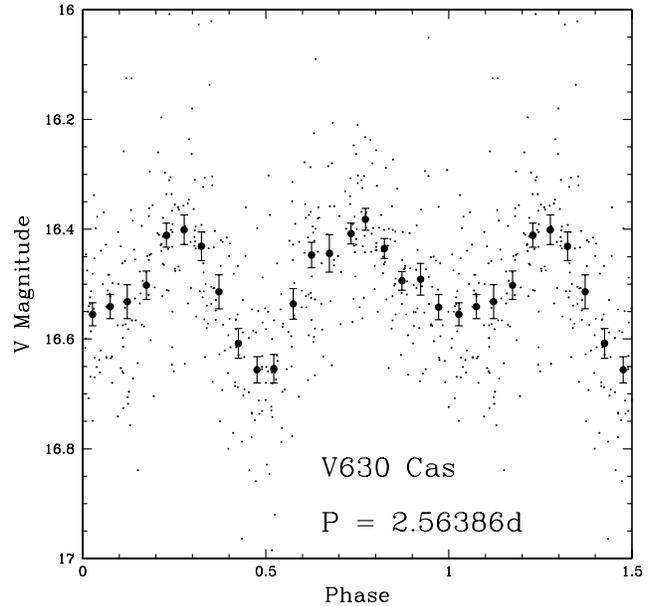} 
\caption{The folded post-outburst $V$-band light curve (small dots).  The
large filled circles are averages over phase bins 0.05 phase units wide.
}
\label{photofold}
\end{figure}

We have folded the post-outburst data on the spectroscopic ephemeris.
Before performing this folding, the post-outburst data for each year
were normalised to a magnitude of 16.5 to remove the long multi-year
trend seen in the post-outburst data of Fig.\ \ref{robo}.  This decreases the
scatter in the folded light curve but does not correct for the
3-9 month changes seen in Fig.\ \ref{robo}.  Fig.\ \ref{photofold}
shows this folded light curve
with superposed averages over phase bins 0.05 phase units wide.
We also constructed a binned light curve by computing the median $V$-magnitude
within each of 20 phase bins using the unnormalised light curve
(excluding the outburst).  The resulting binned light curve is statistically
identical to the binned light curve shown in Fig.\ \ref{photofold}.
Thus we are reasonably confident that the 
binned light curve shown in Fig.\ \ref{photofold} represents the true
long-term averaged light curve.

\subsection{Spectral type of the Secondary}

We used the technique outlined in Marsh, Robinson, \& Wood (1994) to
find the best fitting spectral type of the secondary star and to derive
the spectrum of the accretion disc.  For this we used 26 flux calibrated
spectra from 1994 November, observed with the KPNO 4m.  Each spectrum
was Doppler shifted to zero velocity (using the spectroscopic 
elements in Table \ref{specparm}) before the averaged spectrum was made.
This averaged spectrum, shown as the top spectrum in Fig.\ \ref{spect},
has a signal-to-noise ratio of $\approx 100$ over much of the wavelength
range.  We have 31 observations of 21 different stars (mostly dwarfs) 
observed with the KPNO 4m (13 spectra were from 1993 and 18 spectra were
from 1994).  
The restframe spectrum  and the template spectra were all normalised to 
unity at 5500~\AA.  Then each template was 
scaled by a weight factor $w$, and subtracted from the rest frame
spectrum.  
The scatter in the difference spectrum is measured by
computing the rms from a fourth-order polynomial
fit to the 676 points in the wavelength intervals 5200-5800~\AA\
and 5950-6500~\AA.  The template spectrum
which gives the lowest overall rms,
corresponding to the smoothest difference spectrum, is deemed to be the
best match for the spectrum of the secondary star.

\begin{figure}
\vspace{7.6cm}
\includegraphics{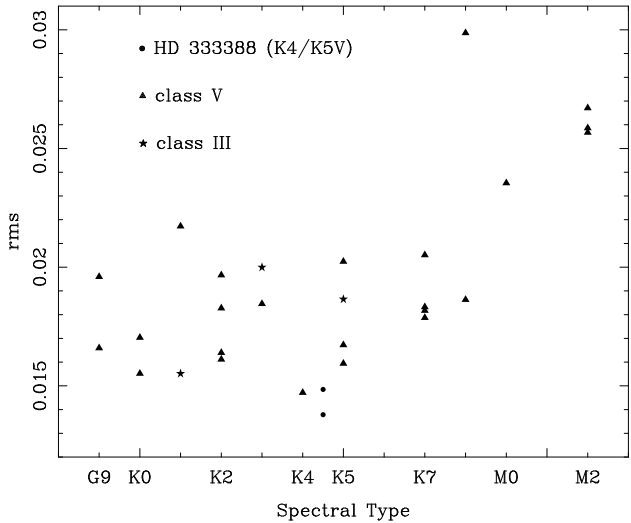} 
\caption{The rms vs.\ spectral type for the 31 template spectra (see text).
The two lowest rms values, and hence the best spectral type match, are
for the two observations of HD 333388, a K4-5 dwarf (filled circles).
The stars are the three luminosity class III templates, and the triangles
are the luminosity class V templates.
}
\label{rms}
\end{figure}

The secondary must be evolved well off the main sequence owing to the
long orbital period.  We have only three
luminosity class III templates (the rest are luminosity class V).
However, our experience with this decomposition technique indicates that
it is not sensitive to the luminosity class of the template at these
low resolutions.  
Fig.\ \ref{rms} shows the rms values as a function of
the spectral type for all of the templates.  The best two matches were 
provided by two different observations of HD 333388, which classified
in the SIMBAD database as K8V.  However, we believe HD 333388 has been
misclassified.  We adopt K4-5V based on a comparison with our other templates,
most notably spectra of 61 Cyg A (a K5V standard) and 61 Cyg B (a
K7V standard).  Apart from HD 333388, the rms  values tend to be
the lowest near K4 and K5.  The K7 and later templates clearly give worse
fits, and the templates earlier than K2 also give relatively poor fits,
although the difference is not as dramatic as for the later templates.
We adopt K4-5 for the spectral type of the secondary in V630 Cas. 

Fig.\ \ref{spect} shows the results of the decomposition.  The top
spectrum is the observed normalised rest frame spectrum.  The bottom
spectrum is the normalised spectrum of HD 333388, scaled by 0.38. 
The spectrum in the middle is the difference between the two.  The 
stellar absorption lines are removed reasonably well, and the only
remaining features are the emission lines, the interstellar
Na D line, and the atmospheric feature near 6870~\AA.  The emission
lines near 4921 and 5015~\AA\ seen in the difference spectrum
are probably due to HeI.  The weak emission
line near 5169~\AA\ might be part of the Fe II
multiplet 42 (the other two lines would be at 
4924 and 5018~\AA, Moore 1972).

We repeated this general prodecure using the higher resolution spectra
in an attempt to measure the rotational velocity of the secondary star.
In this case we have only one template spectrum, that of
HD 3765, a K2V star.  The template spectrum was broadened by various
trial values of $V_{\rm rot}\sin i$ using the standard analytic rotational
broadening kernel (e.g.\ Gray 1992) with a linear
limb darkening coefficient of
0.6 and compared to the spectrum of V630 Cas.  The rotational broadening is
not resolved, and we place an upper limit of $V_{\rm rot}\sin i<40$
km s$^{-1}$.

\subsection{Light curve models}

The binned light curve shown in Fig.\ \ref{photofold} has the characteristics
of the
well-known ellipsoidal variations, namely two maxima of roughly
equal height per orbital cycle and two minima of unequal depth per
orbital cycle.  The character of the ellipsoidal variations is
a strong function of the inclination $i$ and the mass ratio $Q$.  
Therefore, in principle, one can determine the orbital inclination and
mass ratio
by modelling the light curve.  In practice, however, there are difficulties
which can introduce systematic errors into the ellipsoidal modelling.
In the case of V630 Cas, the light from the accretion disc dilutes the
observed ellipsoidal modulations of the secondary star.  
The primary minimum in the light averaged light curve is relatively deep
(it is more than 0.1 mag deeper than the secondary minimum); this larger
than expected depth is almost certainly due to a grazing eclipse of the
star by the disc.  
There is also
a great deal of variability that is not associated with the ellipsoidal
modulations.  However, with the large number of observations the short-term
excursions all seem to ``average out'', and we are reasonably confident that
the mean light curve represents the true ellipsoidal light curve (which is
of course diluted by the disc).

We used the {\sc ELC} 
code \cite{oro00} to model the binned light curve.  We use
specific intensities from the {\sc NextGen} models (Hauschildt et al.\
1999a,b).
The {\sc ELC} 
code can include light from a flared accretion disc, and geometrical
effects due to the disc (i.e.\ eclipses of the secondary star) are accounted
for.  
We have the usual geometrical parameters for the ellipsoidal model, namely
the inclination $i$, the mass ratio $Q$, and the orbital separation $a$.
The main parameters for the secondary star are its filling factor
$f$, its rotational velocity with respect to synchronous $\Omega$, 
its gravity darkening exponent $\beta$, its albedo $A$, and its
mean (or effective) temperature $T_{\rm teff}$.  We assume the star
is exactly  Roche lobe filling and rotating synchronously, hence $f=1$
and $\Omega=1$.  We set $\beta=0.08$ and $A=0.5$, the usual values for
a star with a convective outer envelope.
Finally, 
the spectral type of
the secondary star in V630 Cas is K4-5.  The temperature corresponding to
this depends on the gravity.   For example, a K4V star ($\log g=4.60$)
nominally has
a temperature of 4791~K, whereas 
a K4III star ($\log g=1.7$) nominally
has a temperature of 4113~K, according to the parameters given in
Gray (1992).   A simple two-dimensional interpolation gives
$T_{\rm eff}=4310$~K for a K4.5 star with $\log g=3$ (see below).
The parameters describing the disc are its inner radius $R_{\rm inner}$,
its outer radius $R_{\rm outer}$, the flaring angle of the rim 
$\beta_{\rm rim}$,
the temperature of the inner edge $T_{\rm disc}$, and the power-law exponent
of the disc temperature profile $\xi$, where $T(r)
=T_{\rm disc}(r/r_{\rm inner})^{\xi}$.  Unfortunately, the disc parameters
are not that well constrained for a number of reasons.
For example, we 
have light curves in only one band, so the general slope of the disc
spectral energy distribution is relatively
unconstrained.  In addition, the bright central parts
of the disc are not eclipsed, so it is more difficult to pin down the
disc brightness profile.  
We therefore fixed the inner and outer radii, the flaring angle, and the
power law exponent at reasonable values ($R_{\rm inner}\approx 0.01\,
R_{\odot}$, $R_{\rm outer}=75$\%\ of the primary
Roche radius $R_{RL}$, $\beta_{\rm rim}=2$ degrees,
and $\xi=-0.4$, a value which gives a flatter profile than the nominal
value of $\xi=-3/4$ expected for a steady state disc).  The only
remaining free parameter for the disc in then the temperature at its
inner edge $T_{\rm disc}$.  Values of $T_{\rm disc}$ near 9000~K give
an integrated $V$ band intensity roughly equal to the integrated $V$
band intensity of the star, which is consistent with the observations (e.g.\
the spectral decomposition shown in Fig.\ 
\ref{spect}).  One should bear in mind that the source shows considerable
variability about its mean, hence the ``disc fraction'' measured in
Fig.\ \ref{spect} may be slightly larger or smaller than the true long-term
average.  Finally, although the {\sc ELC} 
code can include the white dwarf,
we omitted the white dwarf for simplicity.  Since the white dwarf is not
eclipsed, its only contribution to the light curve would be constant in
phase.  Hence
there is no real constraint on its temperature and brightness.
Since we have only observations in one bandpass, we can alter slightly
the inner radius and inner temperature of the disc to mimic the
contribution from the white dwarf.  

\begin{figure}
\vspace{6.6cm}
\includegraphics{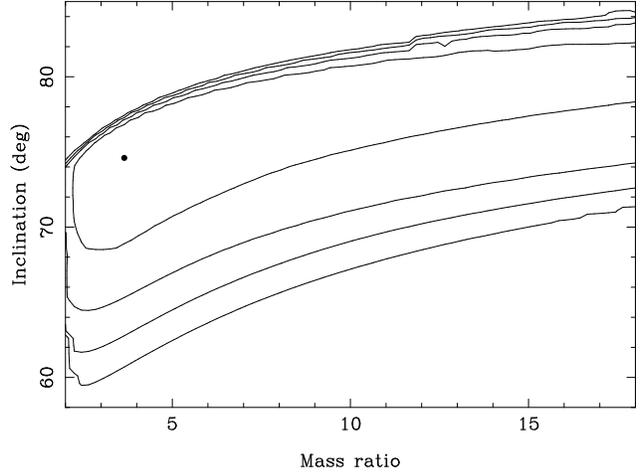} 
\caption{Contours of $\chi^2$ values in the mass ratio-inclination
plane from the light curve fitting.  The filled circle marks the location
of the minimum $\chi^2$ value.  The contours shown correspond to the 
68.3, 95.4, 99.73, and 99.99 per cent confidence levels for
two parameters of interest.
}
\label{plotcont}
\end{figure}

We are left with basically four free parameters:  the inclination $i$, the
mass ratio $Q$, the orbital separation $a$, and the temperature of the
inner edge of the disc $T_{\rm disc}$.  
The two binary observables we have are the radial velocity curve of the 
secondary star and the $V$ band light curve.  To carry out the fitting
of the observables, 
we defined a grid of points
in the $Q$-$i$ plane with steps in $Q$ of 0.1 and steps in $i$ of
$0.25$ degrees.  At each $Q$,$i$ point in the grid, $a$ and $T_{\rm disc}$
are optimised using a ``grid search'' routine adapted from Bevington
(1969) so that the total $\chi^2$ is minimised:
\begin{eqnarray}
\chi^2_{\rm total}&=&\chi^2_{V}+\chi^2_{RV} \nonumber \\
 &=&\sum^{20}_{i=1}{(V_{{\rm obs}}-V_{{\rm mod}})^2_i\over 
\sigma^2_i}+
\sum^{60}_{i=1}{(RV_{{\rm obs}}-RV_{{\rm mod}})^2_i\over 
\sigma^2_i}.
\end{eqnarray}
The errors on the radial velocities were scaled to yield
$\chi^2_{RV}/\nu=1$ for the best-fitting sinusoid.  Fig.\ \ref{plotcont}
shows the contour plot of $\chi^2_{\rm total}$ in the $Q$-$i$ plane.  
The inclination is fairly well constrained to be near 75 degrees, whereas
the mass ratio $Q$ is basically not constrained (we can put independent
constraints on the mass ratio, see Sec.\ \ref{binparm}).
There
is a dramatic and abrupt increase
in the $\chi^2$ values near 80 degrees.  At this inclination and higher 
inclinations substantial eclipses are predicted, and since none are observed
the quality of the fits rapidly get worse.
At inclinations
smaller than about 60 degrees the amplitude of the model is far too small
to match the observations, and the quality of the fits also gets much worse.
The only observational constraint
we have from the spectra is that roughly half of the light is not from
the secondary star.  Our models generally satisfy this constraint.

\begin{figure}
\vspace{6.6cm}
\includegraphics{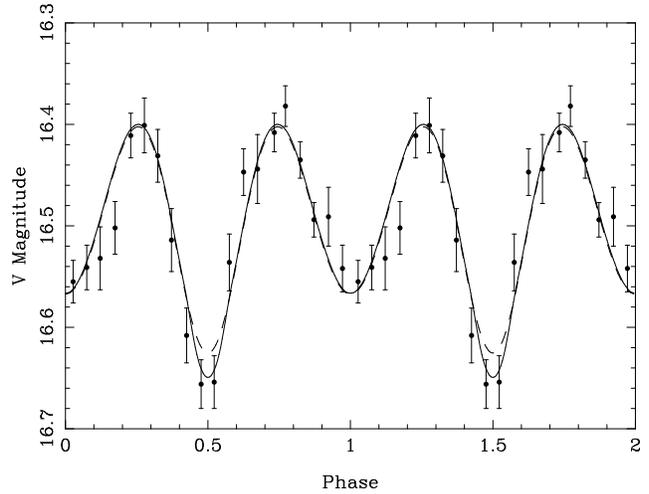} 
\caption{The phase-averaged $V$-band light curve and the ellipsoidal
model (solid line).  The dashed line is a model with light from the disc
but no eclipses.
}
\label{plotlc}
\end{figure}

Fig.\ \ref{plotlc} shows the $V$ light curve and the adopted model
(solid line) with
$Q=5.5$ and $i=74$ degrees (see Sec.\ \ref{binparm}).  The fit is good
($\chi^2=21.39$ for twenty points).  There is a grazing eclipse of the
star by the outer edge of the disc.  The dashed line in
Fig.\ \ref{plotlc} shows what the light curve would look like if this
eclipse is not accounted for (the constant light from the disc is still
included).
The general fitting results are not too sensitive on the assumed disc radius.
We know the disc contributes about half of the light, and if a large
part of the disc is eclipsed then we would expect a $\approx 0.7$ mag
deep eclipse at phase 0.  We do not observe such an eclipse, which means
that the bright central parts of the disc are not eclipsed. 
This in
turn constrains the inclination to be less than about 80 degrees.
In the same vein, the overall amplitude of the light curve is relatively
large ($\approx 0.25$ mag) {\em in the presence of substantial disc light}.  
Thus the inclination cannot be too small.  For most mass ratios, inclinations
smaller than about 65 degrees produce models with amplitudes that are
too small to match the observations.

\section{Astrophysical Consequences}

Having laid out all of the observation of V630 Cas, we now discuss some
of the astrophysical implications of these observations.

\subsection{Mass limits on the components}\label{stripgsec}

The observed radial velocity curve of the secondary star
$K_2$ can be used to place
a straightforward lower limit on the mass of the unseen accreting star
$M_1$:
\begin{eqnarray}
M_1 > f(M) & \equiv &{PK_2^3\over 2\pi G} 
          = {M_1\sin^3 i\over (M_1+M_2)^2} \nonumber \\
&=& 0.624\pm 0.056\,M_{\odot}
\label{massfcn}
\end{eqnarray}
where $M_2$ is the mass of the secondary star and $G$ is the universal 
constant of gravitation.

\begin{figure}
\vspace{6.7cm}
\includegraphics{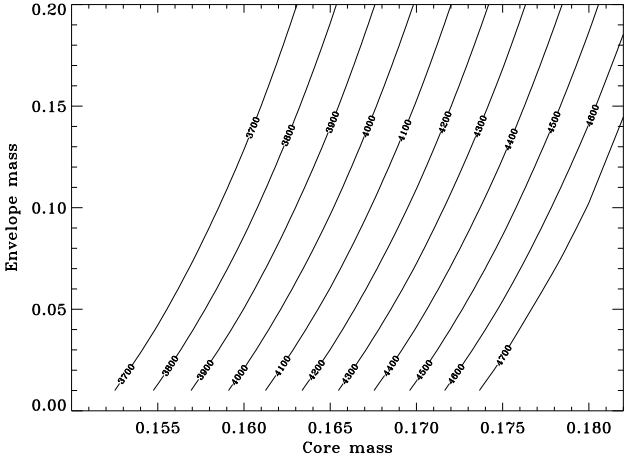} 
\caption{Contours of the effective temperature $T_{\rm eff}$ in the
core mass-envelope plane computed using the relations in
Section \protect\ref{stripgsec}.
}
\label{stripgplot}
\end{figure}

If the secondary star is near the giant branch in the HR diagram, we
can estimate the mass of its core (and therefore a lower limit on
the secondary star $M_2$) using three basic relations obtained
on the evolutionary models of ``stripped giants''.  The luminosity
and radius of the core are strong functions of the core mass
$M_{\rm core}$ (e.g.\ Taam 1983; Webbink, Rappaport,
\& Savonije 1983; also Phinney \& Kulkarni 1994):
\begin{eqnarray}
{L_{\rm core}\over L_{\odot}} &=& 
   \left({M_{\rm core}\over 0.16\,M_{\odot}}\right)^8 
\label{lcore} \\  
{R_{\rm core}\over R_{\odot}} 
   &=& 1.3\left({M_{\rm core}\over 0.16\,M_{\odot}}\right)^5
\label{rcore}
\end{eqnarray}
We also have from Roche geometry and Kepler's Third Law
a relationship between the mean density
of the Roche lobe filling secondary star and the orbital period:
\begin{equation}
\rho={M_2\over 4/3\pi R^3_2}={3\pi\over P^2GR^3_{RL}(Q)}{1\over 1+Q}
\label{den}
\end{equation}
where $Q=M_1/M_2$ and $R_{RL}(Q)$ is the sphere-equivalent radius of the
Roche lobe for unit separation.  
For $2\le Q\le 20$, the quantity $R^{-3}_{RL}(Q)(1+Q)^{-1}$ is
approximately constant 
(it varies between 9.7 and 10.16).  Thus, for our case here, the mean
density of the secondary star is a function only of the orbital period
to a good approximation.  
Therefore the Roche radius is  a function only of the mass of the
secondary star. 
The procedure for finding $M_{\rm core}$
is simple.  For an {\em assumed} core mass $M_{\rm core}$ we know its radius
$R_{\rm core}$ and luminosity $L_{\rm core}$ (eq.\ \ref{lcore},\ref{rcore}).
For an {\em assumed} envelope mass $M_{\rm env}$ we know the total mass
of the secondary $M_2$
and hence its radius $R_2$ using eq.\ (\ref{den}).  
The radius of the secondary star and the luminosity $L_{\rm core}$ can
then be used to find the effective temperature $T_{\rm eff}$.
Fig.\ \ref{stripgplot} shows a contour plot of $T_{\rm eff}$ in the 
$M_{\rm core}$-$M_{\rm env}$ plane for situations where $R_{\rm core}
< R_2$.
The contour lines are nearly
vertical, meaning the core mass $M_{\rm core}$ is tightly constrained
whereas the envelope mass $M_{\rm env}$ is not.  
According to Fig.\ \ref{stripgplot}, $T_{\rm eff}=4300$ corresponds
to $M_{\rm core}\ge 0.165\,M_{\odot}$.  Hence the lower limit on
the secondary star mass is  $M_{2}\ge 0.165\,M_{\odot}$.

\subsection{Binary Parameters}\label{binparm}

We use a Monte Carlo procedure outlined in Orosz \& Wade (1999) to
derive interesting binary parameters and their uncertainties from the
results of the light curve fitting.  Since we have the radial velocity
curve of the secondary star, each point in the mass ratio-inclination plane
corresponds uniquely to a set of binary parameters (e.g.\ component masses,
orbital separation, etc.).  Each point in the mass ratio-inclination plane
also has a certain probability based on the $\chi^2$ value of the light curve
fit.  The Monte Carlo procedure selects parameter sets equally from
different probability intervals and generates probability distributions
for each interesting binary parameter.  Since we don't know the 
mass function (eq.\ \ref{massfcn}) of the accreting star exactly, 20
values of $f(M)$ are drawn randomly from the appropriate
Gaussian distribution at each $Q$-$i$ point in the plane.
We have three additional constraints we add to the Monte Carlo procedure.
The first constraint, $M_1>f(M)=0.624\,M_{\odot}$, is automatically
satisfied.  The second constraint is the one derived from the stripped
giant evolution, namely $M_2>0.165\,M_{\odot}$.  The third constraint is
the upper limit on the rotational velocity, $V_{\rm rot}\sin i<40$
km s$^{-1}$.  For a Roche lobe filling star in synchronous rotation,
\begin{equation}
{V_{\rm rot}\sin i\over K_2}={R_{RL}(Q)\over a}
\left({1+Q\over Q}\right).
\end{equation}
The upper limit of $V_{\rm rot}\sin i<40$ implies $Q>5$ for $K_2=132.9$
km s$^{-1}$.
The results of the Monte Carlo procedure
are shown in Table \ref{astparm}.  

The mass of the white dwarf
is quite close to $1\,M_{\odot}$, while the mass of the secondary star
is only slightly above the lower limit imposed by the core mass:
the hydrogen envelope has a mass of only $\approx 0.01\,M_{\odot}$.  
The predicted rotational velocity of the secondary star (38 km s$^{-1}$)
is just below our upper limit of 40 km s$^{-1}$.  As expected from
the relatively long orbital period, the orbital separation and secondary
star radius are relatively large for a CV ($8.3\,R_{\odot}$ and
$2\,R_{\odot}$, respectively).  The small mass and large radius of
the secondary combine to give a surface gravity of $\log g=3.07$,
considerably smaller than the gravities of late-type main sequence K-stars
($\log g=4.6$, Gray 1992).

\section{Discussion}

The orbital period of 2.56 days is relatively long for a CV.  The
Ritter and Kolb catalog (1998) shows only two CVs with longer periods.
V1017 Sgr is a dwarf nova with a period of 5.714 days (Sekiguchi 1992), 
and MR Vel is supersoft X-ray source with an orbital period
of 3.832903 days (Schmidtke et al.\ 2000).  Two other long period
CVs are GK Per with a period of 1.996803 days (Crampton, Cowley,
\& Fisher 1986) and the old nova X Ser with a period of 1.478 days
(Thorstensen \& Taylor 2000).  
Based on the similarity between V630 Cas, GK Per,
V1017 Sgr, and also BV Cen, Warner (1994) speculated that V630 Cas is a
rare example of a long-period dwarf nova.
Based on correlations between the orbital period $P$ and recurrence time
$T$ and rate of rise $\tau$
\begin{eqnarray}
\log T  & \approx & -0.1+1.7\log P \\
\log\tau & \approx & 0.95 + 0.7\log P,
\label{correl}
\end{eqnarray}
where $P$ is in days, $T$ is in years, and $\tau$ is in days per magnitude,
Warner (1994) predicted an orbital period of $\approx 6$ days for
V630 Cas, more than twice the actual value.  A period of
2.56 days implies $\tau=17.2$ days per mag and $T=3.9$ years.
The observed rate of rise for V630 Cas is about 33 days per mag, although
this is somewhat poorly defined since the two observed outbursts were not
very similar.  The two outbursts were separated by 42 years, and this is
much longer than the 3.9 years derived from eq.\ \ref{correl}.  Indeed,
an inspection of the RoboScope light curve 
(Fig.\ \ref{robo})
suggests that the recurrence
time is more than 8 years, although we cannot rule out the possibility that
a short outburst occurred in a gap of the Roboscope coverage (the RoboScope
light curve shown has 579 observations with the median time 
between observations
1.75 days and the longest time between two observations  153 days).
It seems that the correlations given by eq.\ \ref{correl} have little
predictive power at present, mainly because they are defined by very few
systems at long orbital periods.

\section{Summary}

We have presented extensive photometry and spectroscopy of the dwarf
nova V630 Cas.  The period is relatively long at $P=2.56387\pm 4\times
10^{-5}$ days.  The observed velocity of the secondary star
($K_2=132.9\pm 4.0$ km s$^{-1}$) places a firm lower limit of
$f(M)=0.624\pm 0.056\,M_{\odot}$ for the mass of the white dwarf.
By modelling the ellipsoidal light curve we find an inclination
between 67 and 78 degrees (90 per cent confidence).  Our upper limit on
the observed rotational velocity of the secondary star of
$V_{\rm rot}\sin i< 40$ km s$^{-1}$ implies a mass ratio greater than 5.
Evolutionary considerations give a lower limit of $0.165\,M_{\odot}$ for
the core mass of the secondary star.  All of these constraints yield
component masses of 
$M_1=0.977^{+0.168}_{-0.098}\,M_{\odot}$ for the white dwarf and
$M_2=0.172^{+0.029}_{-0.012}\,M_{\odot}$ for the secondary star
(90 per cent confidence limits).

\section*{Acknowledgments}

It is a pleasure to thank Charles Bailyn for his assistance with the KPNO 
observations.  JRT acknowledges support from the U.S. National Science
Foundation through grants AST-9314787 and AST-9987334.
JAO was supported by a National Young Investigator award from the 
U.S. National Science
Foundation to Charles Bailyn.
This research made use of the SIMBAD data base.

\begin{table}
  \caption{Astrophysical Parameters for V630 Cas}
  \begin{tabular}{lrr}
\hline
parameter & central value & 90 per cent ranges \\
\hline
$i$ (degrees)        & 74.007 & 66.962--78.083 \\
mass ratio           & 5.476 & 4.959--6.105 \\
$M_1$ ($M_{\odot}$)  & $0.976$ & 0.880--1.141 \\
$M_2$ ($M_{\odot}$)  & $0.172$ & 0.165--0.197 \\
total mass ($M_{\odot}$)  & $1.170$  & 1.039--1.325 \\
orbital separation ($R_{\odot}$)  & $8.269$ & 8.024--8.687 \\
$R_2$ ($R_{\odot}$)  & $2.014$ & 1.965--2.110 \\
$L_2$ (bolometric, $L_{\odot}$) & $1.285$ & 1.001--1.839 \\
$\log g_2$ (cgs) & 3.071    & 3.058--3.090 \\
$V_{\rm rot}\sin i$ (km s$^{-1}$) & 38.464 & 37.257--40.000 \\
\hline
  \end{tabular}
\label{astparm}

\medskip
Note:  Additional constraints are $M_2> 0.165\,M_{\odot}$ and
$V_{\rm rot}\sin i< 40$ km s$^{-1}$.
The quote uncertainties are at 90 per cent confidence.
\end{table}

\label{lastpage}

\bsp

\end{document}